\documentclass[prb,twocolumn,showpacs,amsmath,amssymb]{revtex4}
\usepackage{graphicx}% Include figure files
\usepackage[dvips]{color}
\usepackage{bm}% bold math

\def \ed {\end{document}}
\def\Fbox#1{\vskip1ex\hbox to 8.5cm{\hfil\fboxsep0.3cm\fbox{%
  \parbox{8.0cm}{#1}}\hfil}\vskip1ex\noindent}  %%  {TEXT} in BOX
\newcommand{\eq}[1]{(\ref{#1})}%%  requires \eq{label}
\newcommand{\Eq}[1]{Eq.~(\ref{#1})}%%  requires \eq{label}
\newcommand{\Eqs}[1]{Eqs.~(\ref{#1})}%%  requires \eq{label}
\newcommand{\Fig}[1]{Fig.~\ref{#1}}%%  requires \Fef{label}
\let \equiv  \equiv \let\*\cdot \let\~\widetilde \let\^\widehat \let\-\overline

\def\<{\left\langle}    \def\>{\right\rangle}
\def\({\left(}          \def\){\right)}
 \def \[ {\left [} \def \] {\right ]}

\newcommand{\B}[1]{{\bm{#1}}}%% Bold Roman & Greek Lower & Upper Case
\renewcommand{\sb}[1]{_{\text {#1}}}  %% sub-   for lower case
\newcommand{\Sp}[1]{^{^{\text {#1}}}} %% Super- for Upper case
\def\Sb#1{_{\scriptscriptstyle\rm{#1}}}

\definecolor{g-blue}{rgb}{0.83,0.95,1}
\definecolor{g-yellow}{rgb}{1,1,0.7}
\definecolor{g-green}{rgb}{0.9,1,0.9}
\definecolor{green}{rgb}{0,0.6,0}
\definecolor{cyan}{rgb}{0,0.7,0.7}
\definecolor{black}{rgb}{0,0,0}
\definecolor{grey}{rgb}{0.4 ,0.4 ,0.4 }

\def\blue#1{\textcolor{blue}{#1}}
\def\red#1{\textcolor{red}{#1}}
\def\green#1{\textcolor{green}{#1}}

\begin{document}
\preprint{APS/123-QED}

\title{Energy Spectra of Quantum Turbulence: Large-scale Simulation and Modeling}

\author{Narimasa Sasa$^1$, Takuma Kano$^1$, Masahiko Machida$^1$, Victor S. L'vov$^2$,  Oleksii Rudenko$^3$ and Makoto Tsubota$^4$}
\affiliation{
% %	$^1$ CCSE, Japan Atomic Energy Agency, 6-9-3, Higashi-Ueno, Taito, 110-0015, Japan, \\
	$^1$ CCSE, Japan Atomic Energy Agency and CREST(JST), 5-1-5 Kashiwanoha, Kashiwa, Chiba 277-8587, Japan, \\
	$^2$ Department of Chemical Physics, The Weizmann Institute of Science, Rehovot 76100, Israel, \\
	$^3$ Department of Applied Physics, Technische Universiteit Eindhoven, Eindhoven, 5600 MB, The Netherlands, \\
	$^4$ Department of Physics, Osaka City University, Sumiyoshi-ku, Osaka 558-8585, Japan
}%

\begin{abstract}
	In $2048^3$ simulation of quantum turbulence within the  Gross-Pitaevskii equation we demonstrate that the large scale motions have a classical Kolmogorov-1941 energy spectrum $E(k)\propto k^{-5/3}$, followed by an energy accumulation with $E(k)\simeq$ const at \ $k$ about the reciprocal mean intervortex distance. This behavior was predicted by the L'vov-Nazarenko-Rudenko   bottleneck model of gradual eddy-wave crossover [\textit{J. Low Temp. Phys.,} {\bf 153},  140-161 (2008)], further developed in the paper.
\end{abstract}

%\keywords{Suggested keywords}%Use showkeys class option if keyword

\maketitle
\section{\label{s:intro}Introduction}

\emph{\textbf{Hydrodynamic turbulence}} (\textbf{HD})~\cite{Frisch} -- loosely defined as a random behavior of fluids -- remains the most important unsolved problem of classical physics, as was pointed out by Richard Feynman.

\emph{\textbf{Quantum turbulence}} (\textbf{QT}) -- a {trademark} of turbulence in superfluid $^3$He, $^4$He  and in Bose-Einshtein condensates of cold atomic vapors~\cite{VD} -- has added a new twist in the turbulence research shading light on old problems from a new angle.  QT consists of a tangle of quantized vortex lines with a fixed core radius $a_0$ and a finite (quantized) velocity circulation $\kappa= h/M$, where $M$ is the proper atomic mass~\cite{VD}. The superfluid has zero viscosity, and in the \textit{zero-temperature limit}, which is the simplest for theoreticians and reachable for experimentalists~\cite{exp}, the QT's Reynolds number, Re, is infinite. This brings (at least, the zero-temperature) QT to a desired prototype for better insight in the classical HD turbulence.
%This makes QT a more simpler object than the classical HD turbulence, where the concept of vortex tubes or eddies is imprecise and where the viscosity plays an essential role being the only source of the energy dissipation.

The tangle of vortex lines in QT is characterized by a \emph{\textbf{mean intervortex distance}}, $\ell$. For large $R$-scale motions with $R\gg \ell$ the vortex tangles are better understood as bundles  of nearly parallel vortex lines with mean curvature of about $R$~\cite{VD}. For large scales the quantization of vortex lines can be neglected and QT can be considered as the classical one, in which the energy density in the $k$-space, $E(k)$, is given by the celebrated Kolmogorov-1941 (K41) law~\cite{K41}:
\begin{eqnarray} %%
	\label{K41} %
	E\Sb{K41}(k) &=& C\,{\varepsilon^{2/3}}{k^{-5/3}}\,,\\
	E(\B r) &\equiv& {\< |\B u(\B r)|^2\>}/{2} =\!\!\int\!\! E(k) dk \,,  \nonumber%%
\end{eqnarray}
confirmed experimentally and numerically~\cite{Frisch}. Here $C\sim 1$,  $\varepsilon$ is the energy flux over scales, and $E(\B r)$  is the  energy density of turbulent velocity fluctuations per unit mass.
\emph{\textbf{Kelvin waves}} (\textbf{KW}s) are helix-like deformations of vortex lines with wavelength $\lambda$: $a_0\!< \!\lambda\!< \!\ell$. Interactions of KWs on the same vortex line, but with different $k\sim \lambda^{-1}$ lead to the turbulent energy transfer toward large $k$.  This idea (Svistunov \cite{Svistunov95})   was developed and confirmed theoretically and numerically by Vinen~{\it et al.}~\cite{Vinen03},   Kozik--Svistunov (KS)~\cite{KS} and L'vov-Nazarenko (LN)~\cite{LN}. Two versions of KW spectrum where  suggested in Refs.~[\onlinecite{KS,LN}]:
\begin{subequations}
\label{KWS}
  \begin{eqnarray}
    \label{KS} %%
	 E\Sb{KS}(k)&=&C\Sb{KS} \Lambda \,  \varepsilon^{1/5}\kappa^{7/5}\, \ell^{-8/5}k^{-7/5}, \ ~~~~~ \mbox{KS,} \\ % %
	\label{LN}
	 E\Sb{LN}(k)&=&C\Sb {LN}\Lambda \,   \varepsilon^{1/3} \kappa\,  \Psi^{-2/3}\ell^{-4/3}k^{-5/3},\  \  \mbox{LN}, \\
	 \nonumber
	 && \Lambda\equiv \ln (\ell/a_0)\,,
  \end{eqnarray}
\end{subequations}
Here $C\Sb{KS}\gg 1$, $C\Sb{LN}\simeq 1/ \pi $ (Ref.~\onlinecite{LLPN-11}), and $\Psi  < 1 $  characterizes the ratio of $\ell$ to the large-scale modulation of the vortex lines. Parameter $\Lambda\simeq 12 \div 15$ in typical experiments in $^3$He and $^4$He~\cite{exp}. The choice between \Eqs{KWS} is  under intensive  debates~\cite{LN-debate1,LN-debate2,LN-debate3,KS-debate}, which, however, has no principle effect on issues discussed in this paper.

\begin{figure*}
     \includegraphics[width=18 cm]{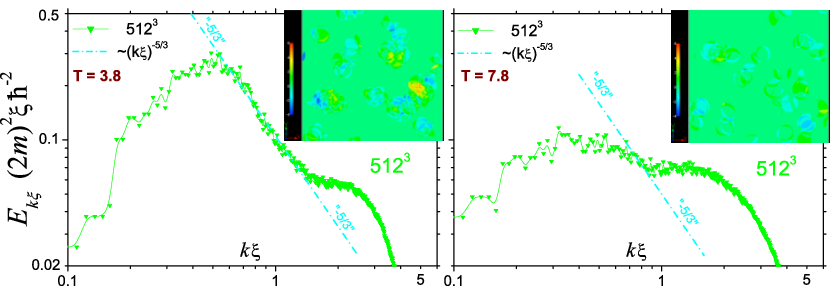}
     \caption{\label{f:1} Color online. The  incompressible  energy spectra   at time T=3.8. Right: the same as the left ones but at later time T=7.8. The color insets show vorticity 2D slice maps (see text for the definition). The grid size is $512^3$ and $\Lambda \simeq 1.5$. }
\end{figure*}

\emph{\textbf{The nature of energy transfer and energy spectrum}} is under intensive debates, too.
Considering the inertial (Re$\rightarrow\infty$) energy transfer at the crossover scale $k\sim \ell^{-1}$, L'vov-Nazarenko-Rudenko (LNR) pointed out~\cite{LNR1} that for $k\sim \ell^{-1}$ and $\Lambda\gg 1$ the KWs have much larger energy~\eq{KWS}  than the HD energy~\eq{K41} at the same energy flux $\varepsilon$.  As the result LNR predicted a bottleneck energy accumulation around  $k\sim \ell^{-1}$. On contrary, KS suggested~\cite{Kozik08} an alternative scenario due to possible dominance of vortex-reconnections in the energy transfer at  $k\sim \ell^{-1}$  without any energy stagnation. In Ref.~[\onlinecite{LNR2}] LNR predicted two thermal-equilibrium regions between the HD\,\eq{K41} and KW\,\eq{KWS} energy-flux spectra: with equipartition of the HD energy, $E(k)\propto k^2$, followed by equipartition of KW energy, $E(k) \simeq $const.

The \emph{\textbf{direct numerical simulations}} \textbf{(DNS)} of QT mostly use the  Gross-Pitaevskii equation (GPE)~\cite{GP}, which in  dimensionless  form is given by %\red{dimensionlessly}:
\begin{equation}
	\label{GPE}
	2\, i \partial \psi/\partial t +  \nabla^2 \psi = g \vert \psi \vert^2 \psi\ .
\end{equation}
The macroscopic wave function $\psi(\B r,t)$ plays a role of the complex order parameter, and  $g$ is the coupling constant. The transformation $\psi = \sqrt{\rho}\,e^{i\theta}$ maps \Eq{GPE} to the Euler equation for ideal compressible fluid of density $\rho$ and velocity $\B u=\B \nabla \theta$, and an extra quantum pressure term.

The numerical study of QT by GPE\,\eq{GPE} has been reported in a few papers so far. Nore {\it et al.}~\cite{Nore97} solved the GPE with resolutions up to 512$^3$ and observed that as the quantized vortices became tangled, the incompressible kinetic energy spectra seemed to obey the K41 law~\eq{K41} for a short period of time, but eventually deviated from it. Kobayashi and Tsubota~\cite{KT05} solved the GPE on $256^3$ grid with an extra dissipation term at small scales, and showed the K41 law~\eq{K41} more clearly. Yepez {\it et al.}~\cite{Yepez09} simulated the GPE on grids up to 5760$^3$ by using a unitary quantum lattice gas algorithm. They also found a spectrum $E(k)\propto k^{-5/3}$  and interpreted it as the K41 law~\eq{K41} of HD turbulence. However, due to the choice of initial conditions, their simulation should correspond to the pure KW region $k>\ell^{-1}$ (thus supporting the LN-spectrum~\eq{LN} of KWs).

\begin{figure*}
	\includegraphics[width=9.1 cm]{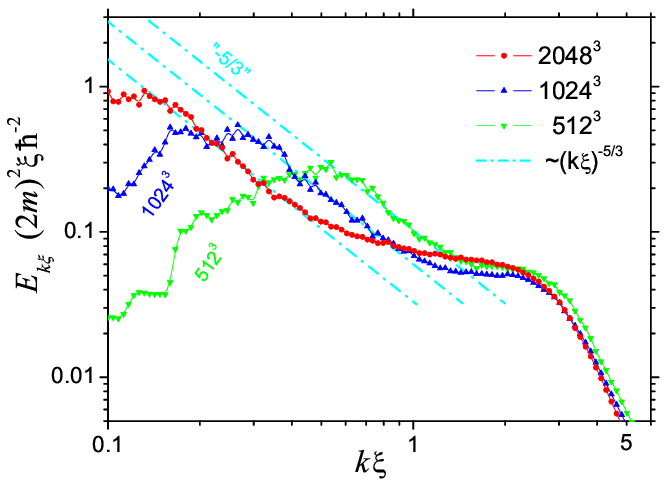}
	\includegraphics[width=8.5  cm]{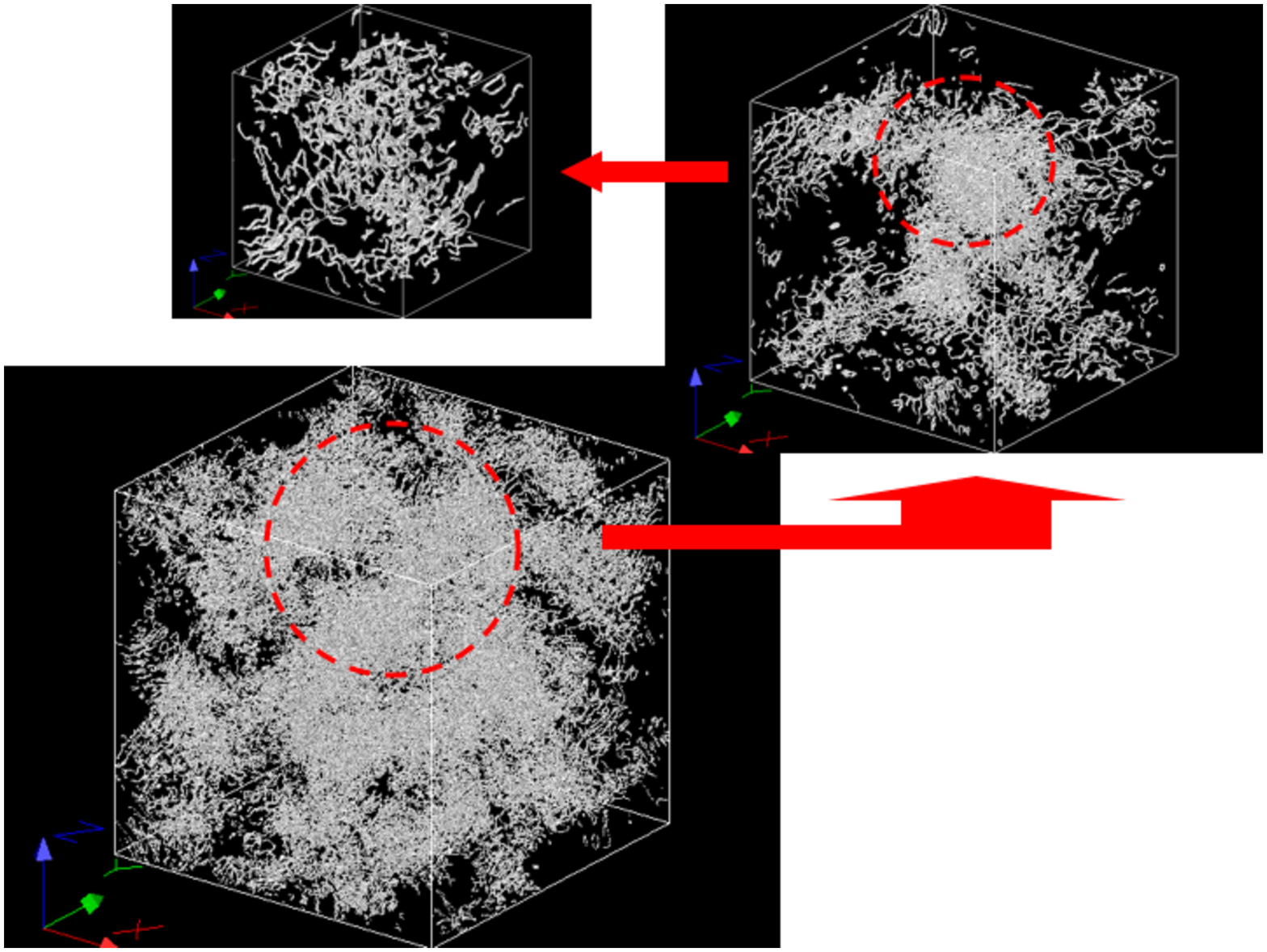}
	
	\caption{\label{f:2}
	  Color online. % % DNS results and LNR-model predictions for the  \red{incompressible} energy spectra in QT.
	  \textbf{Left}: Simulation results of the  incompressible  energy spectra $E(k\xi)$ normalized by ${ \hbar^2/ (4 m^2\xi)}$.  Symbols: $2048^3$ ($\red{-\bullet-}$), $1024^3$ ($\blue{-\blacktriangle -}$), $512^3$ ($\green{-\blacktriangledown -}$).  $\Lambda$ varies from    $\Lambda\simeq 1.5$ (slightly depending on time) for $512^3$  to
 $\Lambda\simeq 2.2$ for $2048^3$.
   \textbf{Left}:  Dot-dashed (cyan) line: K41 ``$-5/3$" scaling.
 A snapshot of vortex lines at the fully-developed turbulent state of $2048^3$ demonstrates the  self-similarity of the bundle-vortex structure (see   the dotted circles  representing the  zoom  regions which vortex distributions are shown subsequently), typical for  fully developed  turbulence.	 }
\end{figure*}

\emph{\textbf{In the present paper}}, we solved  the GPE  on the grids up to  $2048^3$ by parallelizing the simulation code on the Earth Simulator~\cite{ES}. In contrast to Ref.~[\onlinecite{Yepez09}] we focused on HD- and crossover-regions, $k \lesssim \ell^{-1}$.\\
\emph{\textbf{First}}, we confirmed the K41-law~\eq{K41} in the HD-region of about two decades long, which is wider than that of any previous work.\\
\emph{\textbf{Second}}, the visualization of vortices clearly shows the bundle-like structure, which has never been confirmed in GPE simulations on smaller  grids.\\
\emph{\textbf{Third}}, we discovered a plateau in the crossover region, $k\ell\gtrsim 2\pi$, further explained as the KW's energy equipartition in the framework of the LNR's bottleneck model~\cite{LNR2}, which is revised here to account for the recently predicted~\cite{LN} and numerically observed~\cite{Yepez09} LN spectrum~\eq{LN} of KWs.

We consider this correspondence as a   support in favor of LNR bottleneck theory,  understanding, nevertheless,  that interpretation of numerical (or experimental) data with the help of a theoretical model on the  edge of its applicability (here $\Lambda\sim 1$) is often problematic, being a question of experience, physical intuition and taste.
Currently we cannot fully exclude the alternative KS-scenario~\cite{Kozik08}, even though it gives no energy stagnation  for $\Lambda\gg 1$. More theoretical studies, numerical and laboratory experiments are required to fully  understand the vortex dynamics in   the crossover region of scales.

\section{\label{s:DNS}Numerical procedure and results}

In DNS  we   follow techniques~\cite{KT05} but extend the maximum computational grid size from $256^3$ up to $2048^3$. The initial state is prepared by distributing random numbers created inside a range from $-N\pi \alpha$ to $N\pi \alpha$ into the phase $\theta ({\bf r})$ on selected points $M^3$ $(M\ll N)$ and interpolating them to make a smooth velocity field on all grid points. Here $N$ is the total number of grid points and $\alpha$ is a control parameter for the initial energy input. Also, following~\cite{KT05} we add to the GPE an effective artificial energy damping for small-scale motions by replacing in the Fourier transform of GPE  $i \rightarrow i+1$ for   $k_x,\ k_y,\ k_z > 2\pi / \xi$,    where   $\xi \simeq  a_0$ is the condensate coherence length.

GPE conserves the total number of particles and the total energy (Hamiltonian) of the system~\cite{GP}. We decompose~\cite{Nore97,PRA} the total energy density   into the internal, $E\sb{int}\equiv g(\rho-1)^2/4$, the quantum, $E\sb{qnt}\equiv |\B \nabla \sqrt{\rho}|^2/2$, and the kinetic, $E\equiv \rho |\B u|^2/2$, energy densities. The kinetic energy is decomposed into  compressible  and incompressible components, both of which are monitored. Two typical spectra of the  incompressible  component are plotted in Fig.~\ref{f:1} with corresponding vortex distributions. The plot illustrates a $512^3$ run at times $3.8$ and  $7.8$ in the left and right panels, respectively \footnote{The time is normalized by $2m{\xi^2}/{\hbar}$, and distance -- by $\xi$.}. The time evolution of the equation is calculated by a symplectic integral method, and a typical pseudo-spectral method is employed for the calculation of the kinetic energy term. The method is a standard one, which is known to guarantee sufficiently high accuracy for hydrodynamics simulations. In  Fig.~\ref{f:1}, left,  one finds that the major part of the energy spectrum  fits the K41 law~\eq{K41} like in~\cite{KT05}, but with the large inertial interval.

As expected, we also observed tangled vortex bundles clearly demonstrated in the insets of Fig.~\ref{f:1} showing a $x$-$y$ 2D slice of the polarization field's color map, which is defined by summing up vortices ($\pm 1$) inside plaquettes lying within a constant radius ($ = 32\Delta x $) from a grid point. On the other hand,  Fig.~\ref{f:1} is a typical example of a considerably decayed state, in which the main feature is rather small vortex rings distributed almost equally inside the simulation cubic region.

 An important observation (Fig.~\ref{f:1}) is a plateau-like region for $k\xi \gtrsim 1.5 $ -- a definite pile-up over the K41 spectrum -- a clear manifestation of the energy stagnation.

The main numerical result of the present paper is Fig.~\ref{f:2}. The left panel shows an inter-comparison of the  incompressible  kinetic energy spectrum  $E(k)$ among $512^3$, $1024^3$ and $2048^3$  simulations. The  K41  scaling~\eq{K41} (shown as (cyan) dash-dotted lines) is extended to lower $k$ range with the grid-size increase. This is the first clear demonstration of the classical K41 scaling characteristic for the normal-fluid turbulence but maintained in the large-scale range of the superfluid turbulence. The visible extend  of the K41 scaling on $2048^3$ grid is much larger than that in all previous simulations.

The right panel of \Fig{f:2} displays self-similar large structures of tangled vortices in the fully turbulent state: the large-scale vortex bundles in the maximum size, $2048^3$, and smaller self-similar tangled structures inside this cubic region in the subsequent insets.

Before discussion of these results we will revise shortly in the next Section the  LNR model of the bottleneck crossover~\cite{LNR2} to accont for recently predicted LN spectrum of Kelvin waves~\cite{LN}.

% %/Applications/texmakerx.app/Contents/MacOS/texmakerx

 \section{\label{s:LNR} LNR model of the bottleneck crossover}
To find theoretically   $E(k)$ we, following LNR\cite{LNR2},  approximate the superfluid motions as a mixture of ``pure" HD- and KW-motions with the spectra $E\Sp{HD}\!(k)\equiv  g(k \ell)E(k)$ and  $E\Sp{KW}\!(k)\equiv [1-  g(k\ell)]E(k)$.  Here $ g(k \ell)$ is the ``blending" function, which  was found in Ref.~[\onlinecite{LNR2}] by calculation of energies of correlated and uncorrelated motions produced by a system of $\ell$-spaced wavy vortex lines:\\
$\displaystyle
  \label{g}
    g(x)= g_0[0.32 \ln (\Lambda+7.5)x]\,, \  %%
    g_0(x)=\Big[1+\frac{x^2\, \exp (x)}{4\pi (1+x)} \Big]^{-1}$.

\begin{figure} %\vskip .5 cm
	\includegraphics[width=8.5cm]{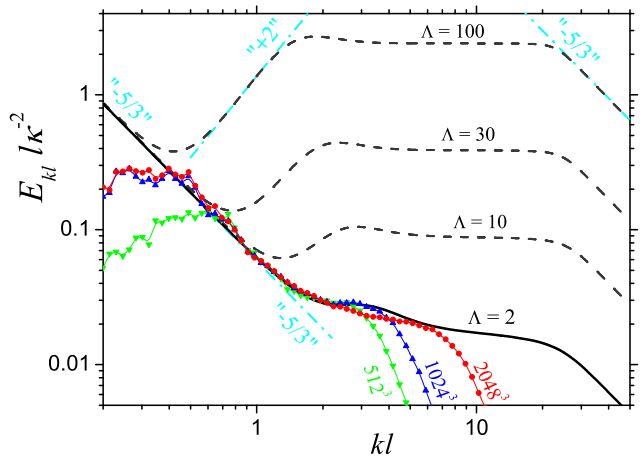}
	\caption{\label{f:3}
	  Color online.
 	Incompressible  energy spectra plotted \textit{vs}. $k\ell$ and normalized by $\kappa^2/\ell$. The simulation results  (the same symbols) and the LNR model for $\Lambda=10,30, 100$ [dashed  (black) curves] are brought together to the theoretical [solid (black)] curve with $\Lambda=2$  by superposing the K41 (both simulations and model) and plateau regions (only for simulations). Dot-dashed (cyan) lines show different scaling asymptotics.}
 \end{figure}

    The total energy flux, $\varepsilon_k$, also consisting of HD and KW contributions~\cite{LNR2}, is modeled by dimensional reasoning in the differential approximation.  Hence, for $k \to 0$ the energy flux is purely HD and thus $\varepsilon_k \propto k^{-2} \sqrt {E\Sp{HD}} d E\Sp{HD}\!\!/ d k$. From the other side, for $k \to \infty$ the energy flux is purely KW and thus $\varepsilon_k \propto [E\Sp{KW}]^2  d E\Sp{KW}\!\!/ d k$. Important that in contrast to the Ref.~[\onlinecite{LNR2}], where the physically irrelevant KS-spectrum  of KWs~\eq{KS} was used, we employ here the proper LN-spectrum~\eq{LN} that accounts for large-scale vortex-line modulations with short KWs~\cite{LN}. The full equation for the total energy flux reads:
\begin{eqnarray}\nonumber
	-\Big\{ \frac 18 \sqrt{k^{11} g(k\ell){  E}(k)} + \frac{3}{5}\frac{\big\{ \Psi k^3 k_*\,\ell^2  [1-g(k\ell) E(k)]\big\}  ^2}{\big ( C\Sb{LN} \Lambda\, \kappa \big)^3 }\Big \}
	\\ \label{mod}
	\times \frac{d}{d k}\Big\{ {  E}(k)\Big[ \frac{g (k\ell )}{k^2}+ \frac{1-g (k\ell)}{k_*^2}\Big] \Big\} =  \varepsilon_k \ . ~~~~~~~~~~~~~~
\end{eqnarray}
Here $E\Sp{HD}\! (k_*) \! =\! E\Sp{KW}\! (k_*)\,, \, \Rightarrow\, k_*\ell \simeq   6.64/  \ln (\Lambda+7.5)$.  In the inertial range  the energy flux is constant $\varepsilon(k)=\varepsilon$. Moreover,  in the system of quantum filaments it is related to the \textit{rms} vorticity $\sqrt{\<|\bm{\omega}|^2\>} \simeq \kappa/\ell^{2}$ via $\<|\bm{\omega}|^2\> = 2\int{k^2 E(k)dk}$ (see Refs.~[\onlinecite{Frisch, VD}]).  This allows to find solutions of \Eq{mod} for different $\Lambda$ as depicted in the \Fig{f:3}  by (black) dashed and solid curves~\footnote{
	For the sake of better comparison we replotted the simulation data bringing them all together to the LNR model curve with $\Lambda=2$ {by superposing the K41 and plateau regions}. It is achieved by fitting the mean inter-vortex distance, $\ell$, which is greed-size dependent: the computation of $\ell$ is approved a-posteriori only if $\ell \gg a_0$; in our case, $\Lambda\sim 1$,  $\ell$ may be considered as a fitting parameter. }.

Four distinct scaling regions are evident ($\Lambda\gg 1$):

I. $k\ell\ll 1$:  $E(k)$  and $\varepsilon_k$ are dominated by the ``pure'' HD contributions, and the K41 law~\eq{K41} is revealed.

II. $k\ell\gg 1$: $E(k)$  and  $\varepsilon_k$ are dominated by the ``pure'' KW contributions, and one observes the LN spectrum~\eq{LN} of KWs with a constant energy flux.

III. $k\lesssim k_*$:  As explained above, for  $\Lambda\gg 1$ the KW turbulence is much less efficient in the energy transfer over scales than its HD counterpart  with the  same energy, which leads to the (HD) energy accumulation with a level $E(k) \approx E\Sp{HD}\!(k) \gg E\Sb{K41}(k)$. For  $k\lesssim k_*$, both $E(k)$  and $\varepsilon_k$ are still dominated by HD contributions, but the energy flux is much smaller than the K41 estimate requires. This is like a flux-free HD system, thus, the thermodynamic equilibrium is expected with the equipartition of energy between the degrees of freedom: 3D-energy spectrum is constant, hence, the 1D energy spectrum  $E\Sp{HD}(k)\propto k^2$. This scaling is observed in Fig.~\ref{f:3}, middle, for  $k\lesssim k_*$. Think of a lake before a dumb, where the water velocity being much smaller than that in the source river does not effect on the water level, which  is practically horizontal. This interpretation  of the energy bottleneck effect as ``incomplete thermalization" (of only high $k$ region) was suggested  by Frisch et. al.~\cite{Fri08}.

IV.  $k\gtrsim k_*$:  Unexpectedly, we observe here almost a $k$-independent 1D-energy spectrum, $E(k)\!\approx$~const, inherent to the thermodynamic equilibrium  of KWs. In the ``pure'' KW system, such a spectrum shows up for $k \gg k_*$. However, in the region IV, the energy of the system is already dominated by the KW contributions, $E(k)\approx E\Sp{KW}\!(k)$, while the energy flux is still dominated by the HD-motions~\cite{LNR2}. Hence, this is almost a flux-free system of KWs, which is indeed found {in the thermodynamic equilibrium  with the 1D energy  equipartition: $E(k)\Sp{KW}\!\!=$ const.}

As one sees from \Fig{f:3}  with the decrease of $\Lambda$ the pile-up becomes less pronounced. For $\Lambda = 2$  the equilibrium HD region (III) almost disappeared, however the equilibrium KW region (IV) is still well pronounced being much less sensitive to the value of $\Lambda$.

\section{\label{s:disc} Discussion and summary}

\subsection{\label{ss:BN}Classical and quantum energy bottleneck effects  }
The bottleneck effect in classical hydrodynamic turbulence is understood traditionally~\cite{VD07,4096} as a hump on a plot of compensated energy spectrum $E(k) k^{5/3}$ in the crossover region between inertial and viscous intervals. This is very general phenomenon reported in   many numerical simulations and experiments
of classical hydrodynamic turbulence. For example, Yeung and Zhou \cite{2}, Gotoh et al \cite{3}, Kaneda et al \cite{4} and Dobler et al \cite{5}
found the bottleneck effect   in their numerical
simulations. Saddoughi and Veeravalli \cite{6}  studied the energy spectrum of atmospheric
turbulence and reported the bottleneck effect. Shen and
Warhaft \cite{7}, Pak et al \cite{8}, She and Jackson\cite{9} and other experimental groups also observed the bottleneck effect in fluid turbulence.  The bottleneck effect has been seen in other forms of turbulence as well, see e.g. Refs. [10-15] in Ref.~[\onlinecite{VD07}].

\begin{figure*}
		\includegraphics[width=8.5 cm]{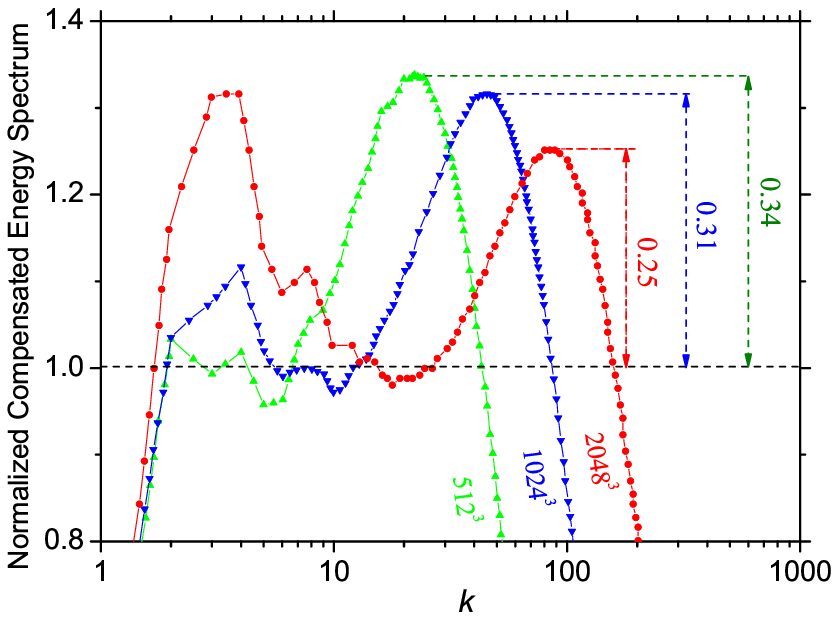}~
		\includegraphics[width=8.5 cm]{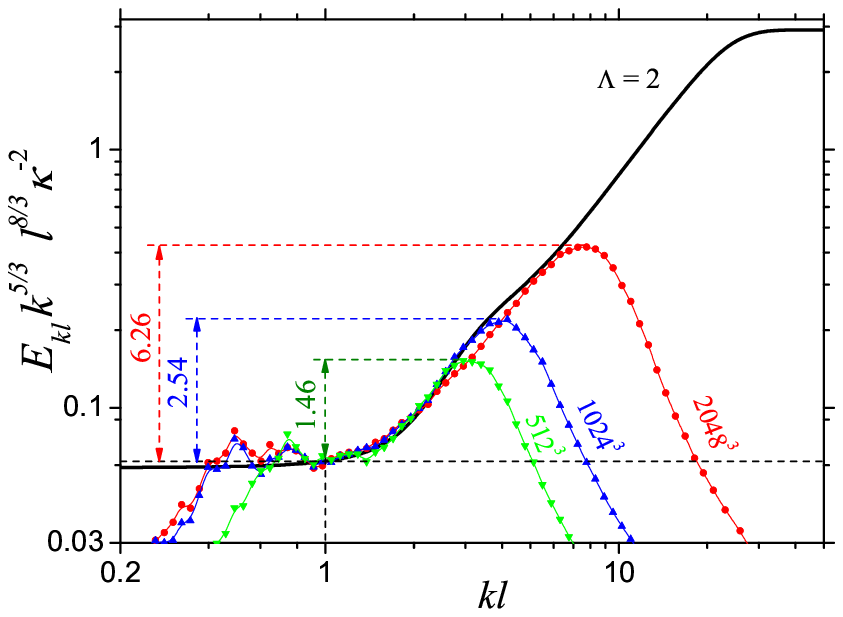}
	 \caption{\label{f:4}
	  Color online. Comparison of the compensated energy spectra, $E(k)\,k^{5/3}$ (the same color-symbol code as in previous figures), and of the bottleneck magnitudes $M\sb{bn}$ (dashed color arrows).
	  \textbf{Left}: DNS data of classical hydrodynamic turbulence~\cite{VD07} (cf. Figure 1 in Ref.~\onlinecite{VD07}): modest values of  $M\sb{bn}\lesssim 1.34$ decrease  with the increase of the resolution.
	\textbf{Right}: Our DNS data of quantum superfluid turbulence: large values of $M\sb{bn}$   increase  with the resolution up to $M\sb{bn} \simeq 6.26$. % (i.e. in 5 times larger that for classical turbulence with the same resolution.
	}
\end{figure*}

To characterize the value of this effect one can introduce a ``bottleneck magnitude" $M\sb{bn}$:  the hump height, normalized by the plateau value of   $E(k) k^{5/3}$ in the inertial interval. For example, in high resolution DNS of the classical hydrodynamic turbulence~\cite{VD07}, shown in Fig.~\ref{f:4}, left, its magnitude   $M\sb{bn}\simeq 0.34$ for 512$^3$ DNS and decreases with the resolution increase: $M\sb{bn}\simeq 0.31$ for $1024^3$ and $M\sb{bn}\simeq 0.25$ for $2048^3$.  {Recent results~\cite{4096} based on the $4096^3$ DNS confirm the statement that the bottleneck magnitude in classical turbulence systematically decreases with the DNS resolution increase (or equivalently, with the Taylor-Reynolds number Re$_\lambda$ growth, and $M\sb{bn} \to \mathrm{Re}_\lambda^{-0.4}$ as Re$_\lambda\to \infty$)}.%%bottleneck magnitude presumably vanishes in the limit  Re$_\lambda\to \infty$ as Re$_\lambda^{-0.4}$).}

Coming to comparison of the bottleneck effects in our modeling  and numerical simulations we should notice  that the LNR model accounts  only for leading  in $\Lambda=\ln(\ell/a_0)$ terms~\cite{LNR1,LNR2}. Moreover, it is based on the differential approximation for the energy flux, which is reasonable for vivid power-like behavior of the energy spectra, which exists only for $\Lambda\gg 1$, Fig.~\ref{f:3}. Therefore, one expects that the LNR model is suitable  for quantitative analysis of experiments in $^3$He and $^4$He, where $\Lambda\simeq 12\div15$, and can only qualitatively describe the simulations presented here with $\Lambda\simeq 2$.

Nevertheless, the simulations clearly demonstrate in Fig.~\ref{f:3} the plateau  that immediately follows the K41-scaling~\eq{K41}, which agrees  with the LNR model prediction for  $\Lambda\simeq 2$ (\Fig{f:3}). The plateau broadens with the grid-size increase towards that of the LNR model curve (the earlier cutoff of the simulation data is due to the artificial dissipation). The resolution of the current simulations does not allow to resolve the KW-scaling~\eq{LN} with constant energy flux as it was done in~\cite{Yepez09}, but the bottleneck is definitely there.

To measure bottleneck  magnitudes in quantum turbulence we re-plotted our data of Fig.~\ref{f:3}, compensating $E(k)$ by K41 prediction, i.e. multiplying by $(k\ell)^{5/3}$, see Fig.~\ref{f:4}, right. One sees large humps with magnitudes $M\sb{bn}$ that increases with the resolution, reaching  $M\sb{bn}\simeq 6.26$ for $2048^3$. Recall that in the classical turbulence $M\sb{bn}$ is much (in about 20 times!) smaller and demonstrates opposite tendency with the resolution.

We concluded  that classical and quantum bottlenecks have completely different nature.  {Small magnitude of the bottleneck in classical turbulence is related to some nonlocality of the energy transfer toward small scales that is slightly suppressed  due to  fast decrease of the turbulence energy in the dissipation range, while in quantum turbulence (at zero temperature) the essential bottleneck effect originates from the strong suppression of the energy flux in the Kelvin wave region}.

{Indeed, Fig.~\ref{f:4}, right, demonstrates a good agreement between the QT DNS data and the LNR model prediction (that accounts for the flux suppression mentioned above)} for $\Lambda\simeq 2$, which improves with increasing the DNS resolution. The cutoff of the spectra for large $k$ is a consequence of limited $k$-space in the simulations. One predicts that with the further increase of the resolution the bottleneck magnitude can reach $M\sb {bn} \simeq 50$ at $\Lambda\simeq 2$ and even much larger values for larger $\Lambda$.

\subsection{\label{ss:disc} Summary }

In the paper we  conclude that the observed essential bottleneck energy accumulation has definitely quantum nature (quantization of {circulation}) and can be completely rationalized within the LNR model of gradual eddy-wave crossover, suggested in Ref.~[\onlinecite{LNR2}].   We consider this model  as a Minimal Model of Quantum Turbulence that   describes homogeneous isotropic turbulence in superfluids with energy pumped at scales much larger than the mean intervortex distance, and reveals reasonable (and even unexpectedly good) agreement with the simulations of Gross-Pitaevskii equation discussed here. The reason is that in the most questionable crossover region, the LNR model predicts a local thermodynamic equilibrium, where the energy spectra are universal and non-sensitive to the details of microscopic mechanisms of interactions, e.g. vortex-reconnections, etc.

 \section*{Acknowlegements}
 We acknowledge  the partial support of a Grants-in Aid  for Scientific Research from JSPS \#  21340104 and from MEXT \#  17071008,  of the Japan Society for the Promotion of Science grant \# S-09147,  of the EU Research Infrastructures under the FP7 Capacities Specific Programme, MICROKELVIN (project \# 228464) and of the U.S. - Israel BSF (grant \# 2008110).

\end{document}